\documentclass[twoside,twocolumn]{article}

\usepackage[sc]{mathpazo} 
\usepackage[T1]{fontenc} 
\usepackage{microtype} 

\usepackage[english]{babel} 

\usepackage[hmarginratio=1:1,top=32mm,columnsep=20pt,left=1in,right=1in]{geometry} 

\usepackage[hang, small,labelfont=bf,up,textfont=it,up]{caption} 
\usepackage{booktabs} 

\usepackage{lettrine} 

\usepackage{enumitem} 
\setlist[itemize]{noitemsep} 

\usepackage{abstract} 

\usepackage{titlesec} 
\renewcommand\thesection{\Roman{section}} 
\renewcommand\thesubsection{\roman{subsection}} 
\titleformat{\section}[block]{\large\scshape\centering}{\thesection.}{1em}{} 
\titleformat{\subsection}[block]{\large}{\thesubsection.}{1em}{} 

\usepackage{fancyhdr} 
\usepackage{titling} 

\usepackage{hyperref} 

\usepackage{hyperref}       
\usepackage{url}            
\usepackage{booktabs}       
\usepackage{amsfonts}       
\usepackage{nicefrac}       
\usepackage{ragged2e}
\usepackage{array}
\usepackage{multirow}
\usepackage{subfig}
\usepackage{amsmath,amssymb}
\usepackage{graphicx}
\usepackage{graphics}
\usepackage{color}
\usepackage{float}
\usepackage{epsf}
\usepackage{caption}
\newcolumntype{L}[1]{>{\raggedright\arraybackslash}p{#1}}
\newcolumntype{K}[1]{>{\centering\arraybackslash}p{#1}}

\pretitle{\begin{center}\Huge\bfseries} 
\posttitle{\end{center}} 
\title{Artery-C \\\LARGE An OMNeT++ Based Discrete Event Simulation Framework for Cellular V2X \\Extended Version\thanks{A  short version of the paper appeared in: A.~Hegde and A.~Festag ``Artery-C --~An OMNeT++ Based Discrete Event Simulation Framework for Cellular V2X'', 23rd International ACM Conference on Modeling, Analysis and Simulation of Wireless and Mobile Systems (MSWiM '20), November 16--20, 2020, Alicante, Spain, DOI:~\href{http://doi.org/10.1145/3416010.3423240}{10.1145/3416010.3423240}}} 

\author{%
\textsc{Anupama Hegde}
\\[1ex]
\normalsize Technische Hochschule Ingolstadt / CARISSMA \\ 
\normalsize \href{mailto:Anupama.Hegde@carissma.eu}{Anupama.Hegde@carissma.eu}
\and
\textsc{Andreas Festag}\thanks{Also with  Fraunhofer Application Center ``Connected Mobility and Infrastructure''} \\[1ex]
\normalsize Technische Hochschule Ingolstadt / CARISSMA \\
\normalsize \href{mailto:Andreas.Festag@thi.de}{Andreas.Festag@thi.de}
}
\date{}
\renewcommand{\maketitlehookd}{%
\begin{abstract}
\noindent 
Cellular Vehicle-to-X (Cellular V2X) is a communication technology 
that aims to facilitate the communication among vehicles and with the roadside infrastructure. Introduced with LTE Release~14, Cellular V2X enables device-to-device communication to support road safety and traffic efficiency applications. We present  Artery-C, a simulation framework for the performance evaluation of Cellular V2X protocols and V2X applications. Our simulator relies on the simulation framework SimuLTE and substantially extends it by implementing control and user planes. Besides the vehicle-to-network communication via the up-/downlink interface, it provides vehicle-to-vehicle and vehicle-infrastructure communication via the  sidelink interface using the managed and the unmanaged mode of Cellular V2X (mode\,3 and~4, respectively). The simulator also implements advanced features of 5G mobile networks, such as variable numerologies. For the transmission of of V2X messages, it adds a non-IP interface. 
Artery-C integrates seamlessly into the simulation framework Artery, which enables the simulation of standardized V2X messages at the facilities layer as well as the coupling to the mobility simulator SUMO. A specific feature of Artery-C is the support of dynamic switching between all modes of Cellular V2X. %
In order to demonstrate the capabilities of Artery-C, we evaluate V2X-based platooning as a representative use case and present results for mode\,3, mode\,4 and mode switching in a highway scenario.
\end{abstract}
}

\begin{document}

\maketitle



\section{Introduction}

\lettrine[nindent=0em,lines=3]{V}2X communication enables the exchange of information among vehicles, roadside infrastructure and other traffic participants. The continuous information exchange supports vehicles to obtain an accurate knowledge of its surrounding environment in order to improve traffic safety and efficiency. Cellular V2X is a mobile network-based communication technology that facilitates the conventional communication between vehicle and network (V2N) to provide backend services and additionally realizes a direct communication among end devices, i.e., vehicle-to-vehicle (V2V), vehicle-to-pedestrian (V2P) and vehicle-to-infrastructure (V2I) communication. The direct communication, also referred to as device-to-device (D2D) communication, allows two physically close end devices to communicate using the sidelink interface (\mbox{LTE~PC-5}). By  sidelink, devices do not have to depend on the cellular access and core network for allocation of radio resources for data transmission~\cite{molina-masegosa2017}. 

\begin{table*}[htb]
\begin{minipage}{1.01\textwidth}
    \caption{Comparison of existing simulation frameworks for Cellular V2X with ARTERY-C} 
    \centering%
    \begin{tabular}{L{2,7cm}L{2,8cm}L{2,5cm}L{2,8cm}L{3,5cm}}%
    \toprule%
    & \textbf{SimuLTE~\cite{virdis2015_simulte_main}} & \textbf{Cellular-VCS~\cite{kuehlmorgen2018}} & \textbf{OpenCV2X~\cite{opencv2x}} & \textbf{Artery-C}\\%
    \midrule%
    Base framework & \texttt{OMNeT++} & \texttt{ns-3} & \texttt{OMNeT++} & \texttt{OMNeT++}\\%
    Protocol stack & LTE & LTE & LTE & LTE, 5G sel. features\\%
    Plane & User & User & User & Control  \& User\\%
    
    Modes & UL, DL, D2D\footnote{Abbreviations: UL = Uplink, DL = Downlink, D2D = Device-to-Device, SL = Sidelink} & UL, DL, SL mode\,3 & UL, DL, SL mode\,4 & UL, DL, SL mode\,3~\&~4\\%
    Mode switching & Cellular\,--\,D2D & No & No & Cellular\,--\,D2D, SL~mode\,3\,--\,4\\%
    Variable numerology & No & No & No & Yes\\%
    V2X applications & IP-based V2I, \newline V2N & IP-based V2I, V2N, V2V~mode\,3 & Non-IP-based V2V~mode\,4 &  IP-based V2I, V2N; Non-IP-based V2V (mode\,3~\&~4)\\%
    Facilities & No & No & Yes, with \texttt{Artery}~\cite{artery} & Yes, with \texttt{Artery}\\
    Open source & Yes & No & Yes & Planned\footnote{We plan to publish the simulation framework under an open source license.}\\%
    \bottomrule%
    \end{tabular}%
    \label{tab:comparison}%
\end{minipage}
\end{table*}%

In comparison to the cellular communication via up- and downlink, the sidelink communication incurs a shorter latency for message transfer, which is a critical aspect for vehicle safety and automation. The sidelink communication enables the transmission of periodic and non-periodic V2X messages as they are defined, amongst other standards, in the European standards for V2X communications, most importantly the Cooperative Awareness Messages (CAM) and Decentralized Environmental Notification Message (DENM). With CAMs, vehicles periodically broadcast their status information such as position, speed, heading to neighboring vehicles with a periodicity of~1 to 10\,Hz. DENMs are triggered in critical safety situations. With the availability of up-/down- and sidelink in Cellular V2X, these  periodic and event-driven messages can be handled across the LTE $U_{u}$ and PC-5 interfaces.

One of the key challenges in modelling vehicular communications for simulation-based performance evaluation is the required integration of a diverse set of components, including vehicle mobility, environmental perception, radio propagation and related effects as well as V2X services and communication protocols. %
The existing \texttt{OMNeT++}-based simulation framework \texttt{Artery}~\cite{artery} -- originally developed for standard-compliant, WLAN-based V2X communication -- provides a clear separation of facilities, application layer and vehicular scenarios, which makes it an ideal base framework for Cellular V2X simulations. %
To model the data plane functionalities of the LTE Radio Access Network (RAN) and Evolved Packet Core (EPC), we have utilized and extended the user plane of the simulation framework \texttt{SimuLTE}~\cite{virdis2015_simulte_main}, implemented the control plane functions, and integrated both user and control plane into \texttt{Artery}, resulting in \texttt{Artery-C}.\footnote{The \texttt{-C} in \texttt{Artery-C} stands for Cellular V2X.}

In \texttt{Artery-C}, the modules for radio resource allocation take into account that a vehicle can be located in the region of cellular coverage and that the allocation process is managed by an LTE base station in a centralized manner (referred to as network-assisted or mode\,3 in 3GPP standards~\cite{ETSI-PHY}).
In case the vehicle cannot remain in the region of cellular coverage, it autonomously configures the radio resources from a pre-defined pool of resources (defined as mode\,4 in 3GPP standards). Hence, \texttt{Artery-C} supports three modes in a common simulation framework, i.e., up-/downlink with the RAN and the EPC, network-assisted sidelink (mode\,3) and out-of-coverage sidelink with distributed resource allocation and management (mode\,4). As an  additional feature to the simultaneous support of all three modes, \texttt{Artery-C} supports dynamic switching among the modes as a part of the simulation scenario. For example, a scenario may involve a change from mode\,3 to 4, when a vehicle moves out-of-coverage and switch back to mode\,3 when network coverage resumes. The same vehicle may communicate simultaneously via the up-/downlink. Hence, the support of mode switching allows modeling more complex scenarios and studying V2X applications under more realistic conditions.

For performance evaluation of Cellular V2X, several simulation frameworks exist (see Tab.~\ref{tab:comparison}). %
A baseline for sidelink in LTE networks has been developed in the \texttt{OMNeT++}-based simulator \texttt{SimuLTE}~\cite{virdis2015_simulte_main} and extended for network assisted device-to-device (D2D)~\cite{virdis2016_simulte_IWSLS}. %
In~\cite{kuehlmorgen2018}, the authors studied Cellular V2X mode\,3 using the \texttt{ns-3} framework. %
The \texttt{OpenCV2X} simulator in~\cite{opencv2x} aims to model and evaluate the performance of sidelink mode\,4 in Cellular V2X. 
However, the existing simulators so far assume only a single resource allocation mode at a given time for a given scenario. %
This leads to a limitation that heterogeneous V2X scenarios with V2V, V2I and V2N cannot be simultaneously studied.

The remainder of the paper is organized as follows: Sec.~\ref{sec:requirements} describes the requirements on a Cellular V2X simulation environment. Sec.~\ref{sec:architecture} explains the implementation design of \texttt{ARTERY-C}. Sec.~\ref{sec:validation} presents simulation results for an example V2X use case using different capabilities of \texttt{ARTERY-C} for the purpose of validation. For the use case, we have chosen V2X-based platooning and study it in a highway scenario for mode\,3, mode\,4 and mode switching. Sec.~\ref{sec:concl} concludes the paper.

\section{Requirements for a Cellular V2X simulation environment}
\label{sec:requirements}

The VANET simulator \texttt{Artery}~\cite{artery} provides a comprehensive framework with a clear separation of the protocol stack and the environment model. %
It allows for a smooth interaction between \texttt{OMNeT++} and \texttt{SUMO}, and adapting the facilities layer to  different access technologies. %
The \texttt{Artery} middleware enables vehicles to use multiple V2X services simultaneously. 

Originally developed for ITS-G5 type of access technologies, the recent version of \texttt{Artery} supports mobile networks, specifically up-/downlink and network-assisted D2D communication~\cite{artery_chapter}. A first approach to extend \texttt{Artery} for sidelink mode\,4 has been addressed in~\cite{opencv2x}. For the development of a comprehensive Cellular-V2X protocol suite with dedicated control and user planes, with a sidelink interface for both mode\,3 and mode\,4, and with support of different V2X application scenarios, we have identified several requirements. 

\subsection{Software-related Requirements}

\textbf{Modularity:} The layers of the protocol suite are defined as modules, which communicate with each other through messages sent across gates. The layers of the user plane -- Packet Data and Convergence Protocol (PDCP), Radio Link Control (RLC), Medium Access Control (MAC) and Physical (PHY) -- are implemented as simple modules and encapsulated into a compound module called Network Interface Card (NIC) as shown in Fig.~\ref{fig:protocolstack}.
The control plane (RRC) is modeled as an independent module, which communicates with the user plane by a message-passing paradigm.

\textbf{Separation between protocol stack and road traffic model:} Traffic models to study different types of V2X scenarios such as V2V, V2I, V2N and V2P are developed using the microscopic road traffic simulator SUMO\footnote{https://sumo.dlr.de (retrieved Sep 10, 2020)}. The Cellular V2X protocol stack is implemented separately in the  \texttt{OMNeT++} simulation framework and  specifically used to study relevant aspects in V2X communication environments, such as resource allocation and scheduling.

\textbf{Message formats for V2X communication services:} 
Currently, the framework \texttt{Artery-C} includes the V2X message types CAM and DENM. The framework aims to support various other message formats for V2X use cases such as infrastructure messages, sensor data  sharing  and maneuver coordination~\cite{3gpp_usecases}.

\subsection{Cellular V2X-specific Requirements}

\textbf{Dedicated sidelink interface:} In order to facilitate an uninterrupted exchange of messages among vehicles, infrastructure and road traffic participants, a dedicated sidelink (PC-5) communication interface is implemented. This interface co-exists with the up- and downlink ($U_{u}$) interfaces. The vehicle communicates with the infrastructure (V2N/V2I) using the up-/downlink and the sidelink is used for V2V, V2I and V2P applications.

\textbf{Sidelink resource allocation modes and dynamic switching:} Both sidelink resource allocation modes,  i.e., mode\,3 and mode\,4, should be supported so that vehicles can exchange messages depending on whether they are located inside or outside the coverage of base station. Dynamic switching between sidelink resource allocation modes causes overhead in terms of mode switching latency~\cite{mode}. 

\textbf{Support of heterogeneous traffic:} 
Depending on the type of application, the framework enables vehicles to simultaneously send both IP-based and non-IP-based data.

\textbf{Resource allocation and scheduling:} Conventional scheduling schemes such as Round Robin (RR), Deficit Round Robin (DRR),  MAXimum Carrier over Interference (MAXCI) and Proportional Fair queuing (PF) are already implemented in the user plane of \texttt{SimuLTE} for uplink and downlink. Following 3GPP standards, the sidelink uses sensing-based semi-persistent scheduling (SB-SPS) for both mode\,3 and mode\,4 (see Sec.~\ref{sec:architecture}).

\subsection{Timing-related Requirements} 

\textbf{Transmit time interval (TTI):} 
The smallest unit step time for the protocol simulations is 1\,ms, which corresponds to the smallest time unit size for resource allocation in LTE. 
Furthermore, 5G-New Radio (NR)(3GPP TS 38.300 V16.1.0) supports a flexible size of resource units in frequency and time, which requires  variable numerologies in the simulator.%

\textbf{Control and user plane latency:} 
Control plane latency is characterized by delays incurred due to communication between the control plane components responsible for registering the end device with the infrastructure and acquiring system information (SI). Correspondingly, the user plane latency is caused by the communication delays between the layers of the user plane. In order to study packet end-to-end latency for various applications, resource allocation during mode switching etc, it is important to understand the impact of both control and user plane latency.%

\textbf{Simulation run time:} 
It refers to the amount of consumed processing time for simulation execution. The simulator allows achieving statistically meaningful results within a reasonable time and commodity computing resources.%

\noindent 
In order to meet the Cellular V2X-specific requirements and realize the software- and timing-related aspects, we have implemented the simulation framework \texttt{Artery-C}.

\section{Implementation design of ARTERY-C}
\label{sec:architecture}

In the \texttt{OMNeT++} framework, the basic implementation unit is called a module.  Modules communicate with each other through event-driven messages. Each module is characterized by a structure defined via
\texttt{.ned} files and a behavior implemented via C++ classes. 
The modules in a network can be of two types -- stationary and dynamic. 
In this section, we present the salient features of our simulation framework \texttt{Artery-C} that has been built as an extension to the user plane implementation in the simulation framework \texttt{SimuLTE}~\cite{virdis2015_simulte_main,virdis2016_simulte_IWSLS}. 

The implementation design of the \texttt{Artery-C} framework consists of two layers. The lower layer covers the protocols of the Cellular V2X access technology. The implementation is aligned with the Cellular V2X protocol stack. It comprises the control plane with RRC and the user plane with PHY, MAC, RLC and PDCP. The layer on top is for generation and reception of V2X messages and represents the facilities of the C-ITS protocol stack.

\begin{figure}[htb]
\centering	
\includegraphics[width=1.0\linewidth]{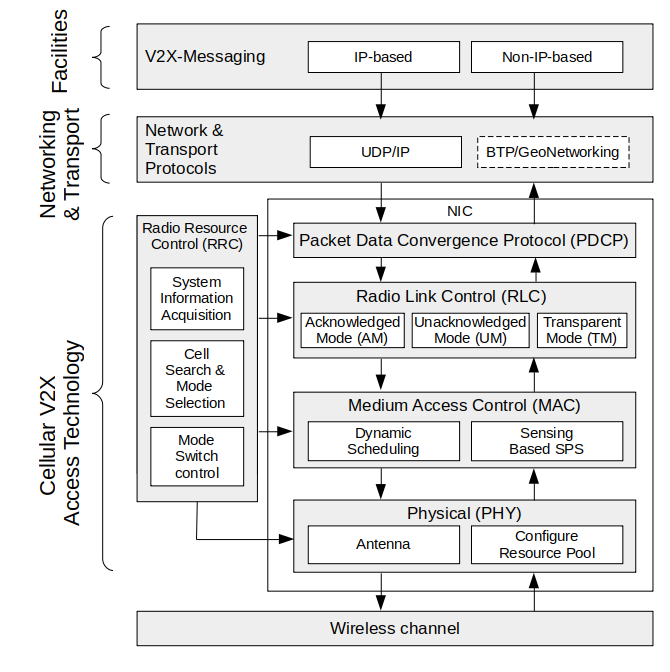}
	\caption{Implementation design of the Cellular V2X stack}
	\label{fig:protocolstack}
\end{figure}

The user plane has two parallel pipelines for IP-based and non-IP-based traffic (Fig.~\ref{fig:protocolstack}) with clearly separated functionalities. The facilities layer realizes the V2X messaging, such as the CAM service modules of \texttt{Artery} to generate non-IP-based periodic messages. For IP-based V2I traffic, \texttt{Artery-C} re-utilizes the modules from \texttt{INET} and \texttt{Artery}. 
The user plane is directly linked with the SUMO traffic model via the \texttt{TRaCI} API in order to continuously track the position of the vehicle and identifies whether it is located in the region of cellular coverage or not. The LTE base stations (eNodeB) and road side units (RSUs) are modelled as stationary modules and vehicles are modelled as dynamic modules.

\subsection{Control Plane -- Radio Resource Control}

\textbf{Cell search and mode selection:}
The control plane/Radio Resource Control (RRC) component is modelled in accordance with the 3GPP Release~15 standards for 5G-NR (TS 38.331, V15.7.0). The RRC is responsible to carry out three primary functionalities, i.e., system information acquisition, cell search and mode selection \& mode switching control. Based on the position updates of the vehicle from the \texttt{SUMO} traffic scenario, the cell search module of the RRC determines the distance between an UE and an LTE base station (eNodeB). If the UE is located inside the communication range of the eNodeB and the received signal strength meets the threshold limits, then mode\,3 is the preferred mode of operation. If an UE lies outside the region of a base station's communication range, then mode\,4 is selected. 

\textbf{System information acquisition:}
The process of system information acquisition involves the exchange of network-related messages between UE and eNodeB that enables the UE to establish a successful connection with the eNodeB and the components of the Evolved Packet Core (EPC).
When an UE recognizes itself to be in the region of cellular coverage, it sends a $RRC_{ConnSetup}$ request to the eNodeB. The eNodeB responds with the appropriate Master Information Block (MIB) and System Information Block (SIB) for different communication interfaces, i.e., UL, DL and SL. Additionally, the eNodeB pre-configures a set of time and frequency resources in the form of SIB that can be used by the UE when it operates in mode\,4. The synchronization-related information~\cite{mode} between UE and eNodeB are exchanged immediately after the establishment of successful connection.

\begin{figure}[h]
\centering	
\includegraphics[width=0.5\linewidth]{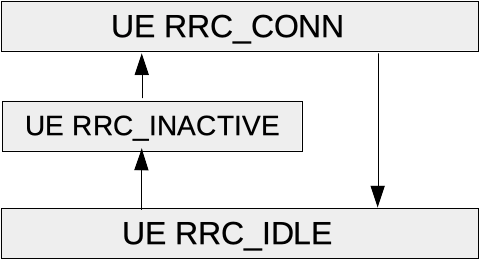}
	\caption{RRC state transition}
	\label{rrc_states}
\end{figure}

\textbf{Mode switching control:} The mode switching control module is responsible to regulate the dynamic switching between mode\,3 and mode\,4. This is based on the availability of network coverage, received signal strength and traffic load. The resource allocation modes directly correlate with the operating states of the RRC - $RRC_{IDLE}$, $RRC_{INACTIVE}$ and $RRC_{CONN}$ and correspond to the three-state finite state machine (FSM) as shown in Fig.~\ref{rrc_states}.

In the $RRC_{CONN}$ state, the UE is connected to the eNodeB and has established its identity with the EPC. The eNodeB is responsible for allocating subchannels and subframes for sidelink communication. When an UE is inside the network coverage but not exchanging any information with the eNodeB, it transits to $RRC_{INACTIVE}$ state. In this state, it is still registered with eNodeB and hence continues to function in mode\,3. When an UE moves outside the base station coverage, the RRC transits to $RRC_{IDLE}$ state and it informs the PHY and MAC component to allocate resources from a pre-configured pool of resources mentioned in appropriate SIBs. The UE is now completely disconnected from the base station.

\subsection{User/Data Plane}

\textbf{Packet data and convergence protocol (PDCP):}
The PDCP is the connecting component between the Cellular V2X access technology and the networking \& transport layer. It processes both IP and non-IP packets. In case of IP traffic, it performs Robust Header Compression (ROHC) and assigns/creates the Connection Identifier (CID) that uniquely identifies, together with the UE ID, a connection in the whole network. When an IP packet arrives at PDCP, a logical connection identifier  (LCID) is attached to it and forwarded to the radio link control (RLC). %
In case of a non-IP-based packet\footnote{In the European C-ITS standards, non-IP packet transport is realized by GeoNetworking (ETSI EN 302 636-4), an ad hoc network protocol based on geographic positions, and BTP , an UDP-like transport protocol (ETSI~EN~302~636-5). Both are implemented in the simulation framework Artery but beyond the scope of this paper.} the PDCP performs ROHC, creates an entry in the non-IP connections table and forwards the packet as a PDU to the RLC. The two pipelines are illustrated in Fig.~\ref{fig:protocolstack}.

The PDCP consists of two separate gates for data input from IP and non-IP traffic. Additionally, it has a separate gate to receive control-related messages from the control plane. For interaction with the RLC, similar to~\cite{virdis2015_simulte_main}, three different gates are connected with the PDCP-RRC module, one for each RLC mode.

\textbf{Radio link control (RLC):}
The RLC operates in three modes - acknowledged (AM), un-acknowledged (UM) and transparent mode (TM). The key functionality of this component is to multiplex and de-multiplex MAC SDUs to/from the MAC. The implementation of the RLC has not been modified much in reference to~\cite{virdis2015_simulte_main}. %
For sidelink broadcast operation, an acknowledgement is not applied and we use  the un-acknowledged mode.

\subsection{Medium Access Control (MAC)}

\begin{figure}[h]
    \centering	
    \includegraphics[width=0.98\linewidth]{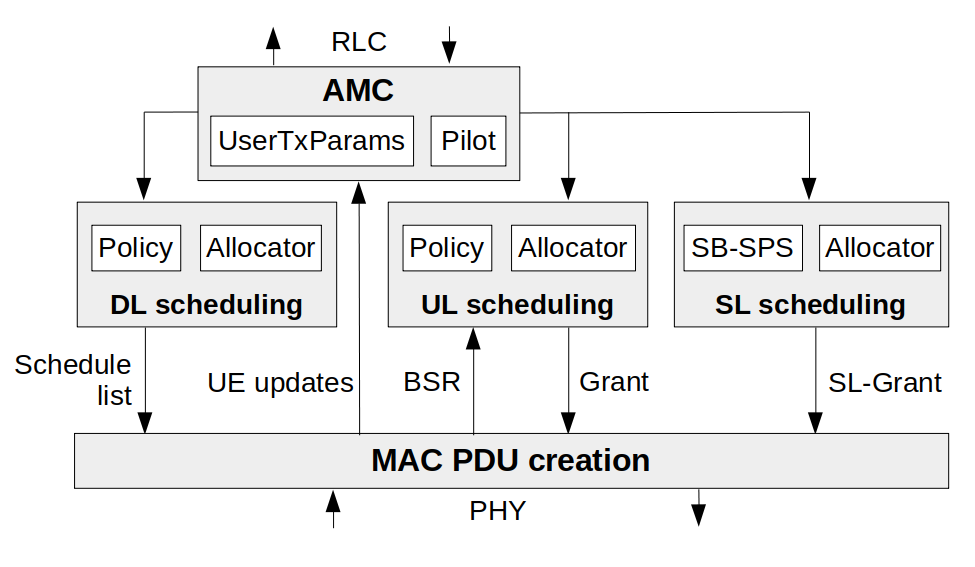}
	\caption{MAC scheduler}
	\label{mac_scheduler}
\end{figure}

The design of the MAC for the UE module has been modified in a way that the sidelink scheduling co-exists with the previous uplink and downlink implementations as done in \texttt{SimuLTE}~\cite{virdis2015_simulte_main,virdis2016_simulte_IWSLS}. %
Additionally, the MAC module has a separate gate to receive control-related information from the control plane and incorporates an additional sub-module called sidelink configuration (SC). %
The sidelink configuration sub-module interacts with the sidelink resource allocation (SRA) submodule in the PHY component and together they support the process of resource configuration and allocation ~\cite{mode}.
A functional block diagram of the MAC scheduling module is depicted in Fig.~\ref{mac_scheduler}. %
The adaptive modulation and coding (AMC) module stores channel status information, which is based on the periodic feedback from the UE via up-/downlink. For sidelink operation, the chosen modulation scheme is fixed as quadrature phase shift keying (QPSK). In addition to the existing scheduling policies, MAXCI, PF and DRR, we have implemented sensing-based semi persistent scheduling (SB-SPS) module for sidelink scheduling. 

\textbf{Mode 3 operation}
The eNodeB has complete knowledge about the registered UEs and the resources utilized by them. In the existing UL and DL implementation, the eNodeB creates a scheduling list and provides the set of available resources in every TTI. When an UE wants to transmit data on the sidelink, based on it's geographical location, it connects to the nearby eNodeB and requests for time and frequency resources. The UE reports to the eNodeB about the size of the data, periodicity and maximum allowed latency based on the type of V2X application. The ``sidelink configuration'' submodule in the MAC of the eNodeB configures the sidelink grant and requests the PHY component to allocate the candidate resource pools (CSRs) and generates a sidelink control information (SCI) message.

On receiving the CSRs from the PHY component, the MAC of the eNodeB updates the resource-related information in the sidelink grant and proceeds with scheduling of resource for sidelink. %

\textbf{Mode~4 operation:}
In mode\,4 operation, the MAC of the UE handles the scheduling of resources independently of the base station. The channel-related cost metrics such as signal-to-noise-ratio (SINR), the reference signal received power (RSRP) and the received signal strength indicator (RSSI) are computed in the PHY component, which play a key role in determining the list of CSRs. On receiving the pool of CSRs from the PHY component, it further handles the SPS information and schedules time-frequency resources for both data and SCI. 

\begin{figure}[h]
    \centering	
    \includegraphics[width=0.7\linewidth]{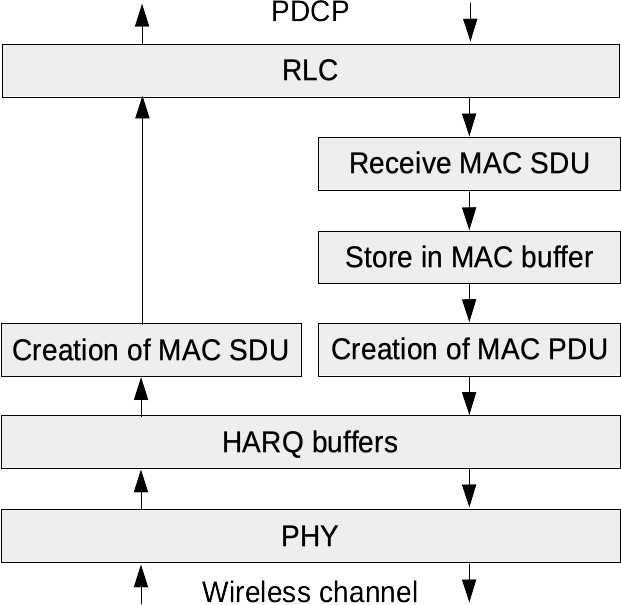}
	\caption{Flow of packets in the MAC component}
	\label{mac_packetflow}
\end{figure}

The flow of packets in MAC buffers is shown in Fig.~\ref{mac_packetflow}.
On obtaining the service data unit (SDU) from the RLC, they are stored inside MAC buffers. %
Based on the scheduling list generated by the scheduler, MAC protocol data units (PDU) are created and stored in HARQ buffers. %
There are separate transmission and reception HARQ buffers to store MAC PDUs that are sent and received. %
The HARQ buffers in eNodeB contain MAC PDU information for each of its connected UEs in both uplink and downlink. %

\subsection{Physical (PHY) component}

\begin{figure}[h]
    \centering	
    \includegraphics[width=0.9\linewidth]{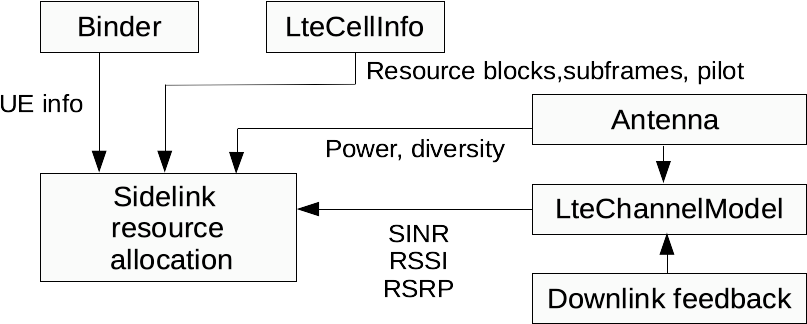}
	\caption{Physical (PHY) component}
	\label{phy_modules}
\end{figure}

The PHY module contains information about antenna power characteristics, standard channel models and cell related information (macro, micro and pico cells). %
The functional components of the PHY module are depicted in Fig.~\ref{phy_modules}. %
In the context of our work, the key functionality of the PHY component is to allocate a set of candidate resource pools (CSRs) for sidelink broadcast communication through SB-SPS. %
The ``sidelink resource allocation'' sub-module is responsible for computing the CSRs by utilizing the cell-related information and obtaining the channel-related parameters from the channel modules. 

\textbf{Sidelink resource allocation (SRA):}
The sidelink control information (SCI) is a 32 bit sequence that is transmitted prior to the transmission of transmit block (TB). In both mode\,3 and mode\,4, the SCI is transmitted over two resource blocks in the same subframe as~TB. %
The TB carries the payload data, which are generated in the facilities layer. 
The time resources are allocated in the form of subframes and the frequency resources are characterized by subchannels, which comprise a group of physical resource blocks (PRBs). The number of subchannels ($N_{subCH}$) and size of subchannels (i.e., number of PRBs per subchannel) ($N_{PRB}$) can vary in a certain range as specified in~\cite{ETSI-PHY}.

The set of subframes belonging to PSSCH pool (mode\,3 \& mode\,4) is denoted by
$[t^{SL}_{0}, t^{SL}_{1}, \cdots t^{SL}_{max}]$. This pool includes all subframes except the subframes where sidelink synchronization signals (SLSS) is transmitted. Synchronization subframes occur periodically at every 160\,ms.

The frequency resource pool consists of a set of subchannels ($N_{subCH}$) comprising of contiguously allocated resource blocks ($N_{PRB}$). The SCI and TB can be transmitted in adjacent or non-adjacent resource blocks of the same subframe. If SCI and TB are transmitted on adjacent resource blocks, the subchannel $m$ comprises a set of contiguous resource blocks calculated as

\begin{equation}
    n_{PRB} = n_{subCHRBStart}+m*n_{subCHsize}+j+\beta
\end{equation}

where $m=0,1, \cdots N_{subCH-1}$ and $j=0,1, \cdots n_{subCHsize}-1$. The starting index of subchannel, $ n_{subCHRBStart}$ is indicated by the higher layer components. Here the value of $\beta$ is~2.

\begin{figure}[htb]
	 \centering
	 \includegraphics[width=0.99\linewidth]{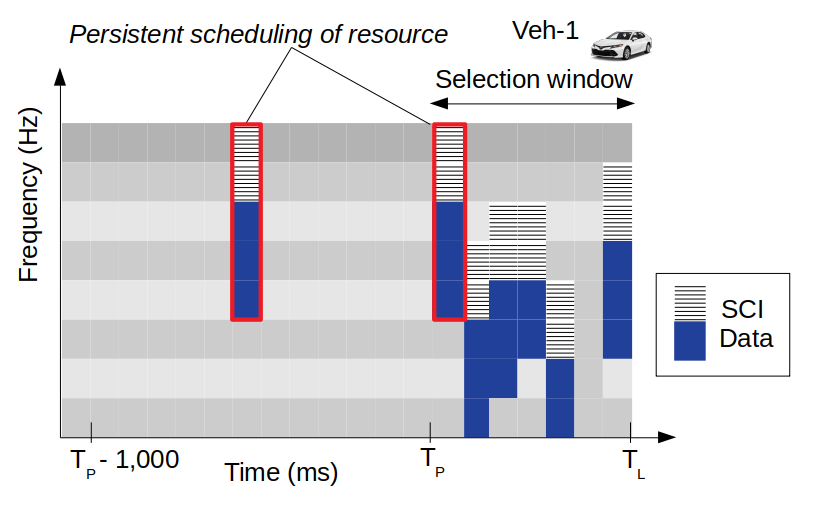}
	 \caption{Sensing-based semi-persistent scheduling (SPS)} 
	 \label{fig:mode4a}
\end{figure}

\textbf{Selection of candidate single-subframe resource (CSR) pool:}
The PHY module determines the CSRs based on sensing-based semi persistent scheduling (SB-SPS). In mode\,3, the procedure is carried out by the eNodeB in a centralized manner and in mode\,4, it is carried out by the UE in a distributed way.
The time interval between the generation of a packet $T_{p}$ and the maximum allowed latency $T_{L}$ is known as the selection window. The TTIs and the resource blocks (RBs) consist of subcarrier groups in a time-frequency grid as depicted in Fig.~\ref{fig:mode4a}. The minimum number of RBs needed to transmit an SCI is two~\cite{ETSI-PHY,book_5G}. The number of RBs for data transmission varies depending on the size of the transmit data block (TB).

In Fig.~\ref{fig:mode4a}, Veh-1 is our desired transmitter UE for which resources have to be allocated. In each TTI of the selection window, Veh-1 identifies a list $L_{1}$ of $M_{total}$ candidate resources that are needed for the transmission of both SCI and TB (adjacent RBs are allocated for SCI and TB in Fig.~\ref{fig:mode4a}).
A candidate single-subframe resource $R_{x,y}$ is a set of contiguous subchannels $L_{subCH}$. The selection window time interval is $[n+T_{P}, n+T_{L}]$, where $T_{P}\leq 4$\,ms and $20 \leq T_{L} \leq 100$\,ms~\cite{ETSI-PHY}. 

From the above list $L_{1}$, Veh-1 discards those resources that are affected by the events below and creates a new list $L_{2}$. %
Next, Veh-1 discards the resource elements in the list $L_{1}$, which are already reserved (persistently scheduled) by another UE in the previous 1,000\,TTIs (highlighted in red color in Fig.~\ref{fig:mode4a}). The UE decides to discard a certain subframe $y$ in the selection window in accordance to Eq.~\ref{eq:csr_selection} where there is an integer $j$ that meets

\begin{equation}
\label{eq:csr_selection}
   y + j * P^{'}_{rsvp-TX} = z + P_{step}* k
\end{equation}

\noindent $z$ is a subframe in the sensing window, the resource reservation interval given by higher layers is \mbox{$P^{'}_{rsvp-TX} = P_{rsvp-TX}*P_{step}/100$}, $j = 0,1, \cdots C_{resel} -1$. If the RSRP value measured over any resource element in the list $L_{1}$ exceeds a given threshold, it indicates that another UE is currently using it for its transmission.%

Note that the number of resources in $L_{2}$ must contain at least $20\,\%$ of resources in the selection window. Otherwise, we increase the RSRP threshold by 3\,dB and iterate again. %
From the list $L_{2}$, Veh-1 ranks the resource elements in increasing order of their RSSI values and identifies the ones with low RSSI values thereby creating a new list $L_{3}$. %
These resource elements are preferred because a low value of RSSI indicates that it has not been used by any other UE during the TTI of our interest. %
From this list $L_{3}$, \mbox{{Veh-1}} autonomously chooses any of the available resources and uses it for transmission. 

\textbf{Mode~3 operation:}
In mode\,3, the subframes and subchannels for transmission of SCI and TB are allocated by the eNodeB. From the standards perspective, Eq.~\eqref{eq:csr_selection} is adapted as

\begin{equation}
    y + j * P^{'}_{SPS} = z + P_{step}* k
\end{equation}

where $ P^{'}_{SPS}$ is the sidelink SPS interval of the corresponding SL SPS configuration given by higher layers, $ P^{'}_{SPS} = P_{SPS}*P_{step}/100$. 

\textbf{Mode~4 operation:}
Utilizing the sidelink grant, the UE monitors the resources utilized by other UEs in the previous 1,000\,ms interval, which is known as sensing. The UE uses a resource re-selection counter $C_{resel}$ whose value is decremented by one every time a packet is successfully transmitted. Once the $C_{resel}$ value reaches zero, the UE has to perform sensing and allocate resources.

\begin{figure*}[htb]
    \centering
    \subfloat[Non-IP, V2V broadcast among platoon members (PM) and platoon head (PH) and IP, V2I unicast data traffic between PH and roadside unit (RSU)]
    {\includegraphics[width=0.49\linewidth]{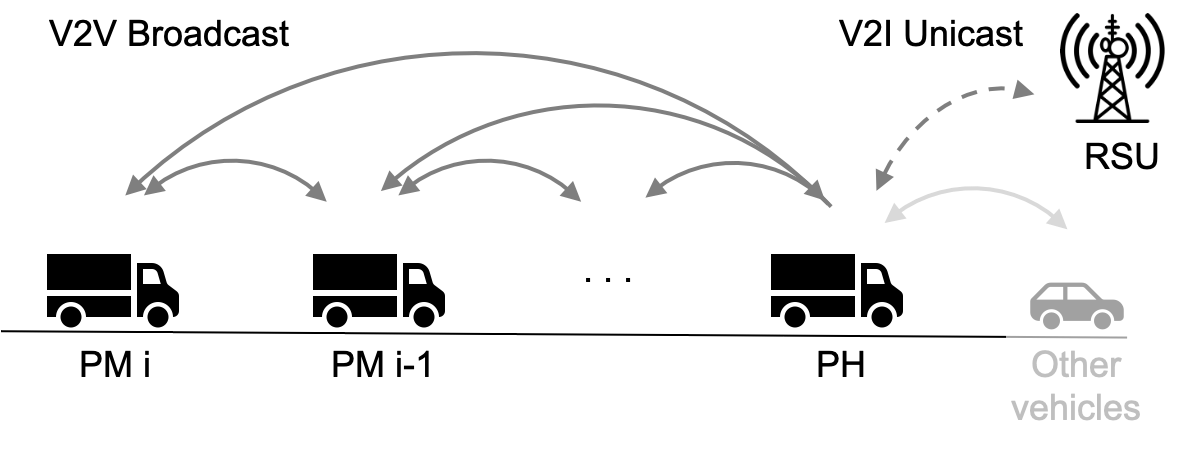}
    \label{subfig:usecase}}
    \hfil
    \subfloat[Scenario with adjacent regions with and without eNodeB coverage]
    {\includegraphics[width=0.49\linewidth]{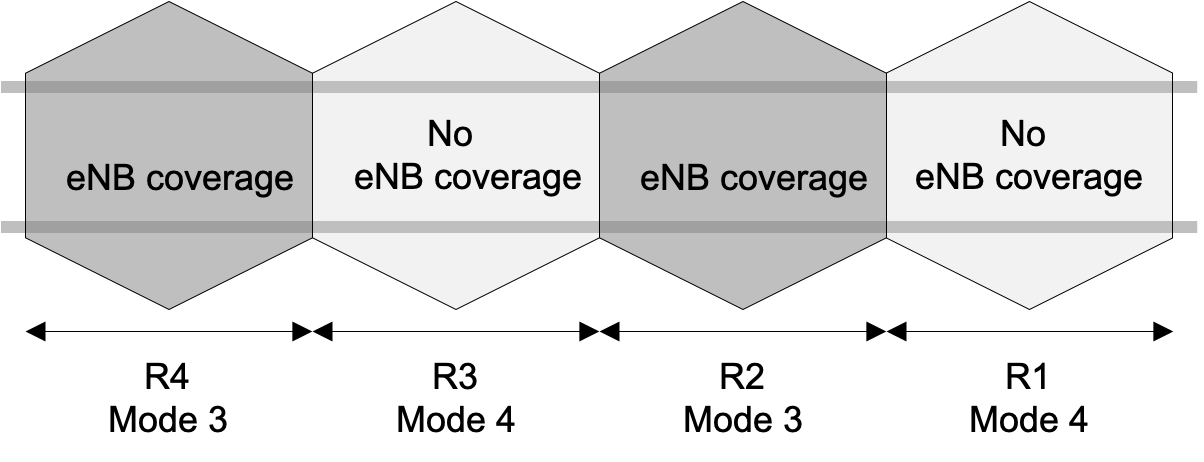}
    \label{subfig:scenario}}
    \caption{Use case V2X-based platooning with mode switching between adjacent regions of cellular coverage and non-coverage} 
    \label{fig:platooning-usecase-scenario}
\end{figure*}

\section{Validation of the \texttt{ARTERY-C} simulator} 
\label{sec:validation}


In order to validate the implementation of the Cellular V2X protocol stack in the simulation framework and to demonstrate its capabilities, we assess the performance of V2X-based platooning~\cite{3gpp_usecases}. %
The implementation of the use case applies sidelink and a simultaneous flow of IP and non-IP based data traffic for V2I and V2V (Fig.~\ref{subfig:usecase}). %
We test the use case in a highway-tunnel scenario with a truck platoon and surrounding vehicles with three variants: \emph{(i)}~full eNodeB coverage, \emph{(ii)}~no eNodeB coverage and \emph{(iii)}~adjacent regions with and without eNodeB coverage. The latter option implies mode switching, see Fig.~\ref{subfig:scenario}. %
The scenario-related simulation parameters are listed in Table~\ref{tab:params}.

\begin{table*}[htb]
    \centering
    \caption{Scenario-related simulation parameters}
    \label{tab:params}
    \begin{tabular}{lcc}
    \toprule
        \textbf{Parameter}  & \multicolumn{2}{c}{\textbf{Road type}}  \\ 
        \cmidrule{2-3}
        &  Highway & Tunnel  \\
        \midrule
        Vehicle speed [km/h] & 100\,--\,130 & 60\,--\,80  \\
        Range of sidelink broadcast [m] & 100 &80 \\
        Cellular coverage regions (Fig.~\ref{subfig:scenario})& R2, R4 & R1, R3 \\
        Types of vehicles & \multicolumn{2}{c}{Cars, trucks} \\
        Pedestrians/slow moving vehicles & \multicolumn{2}{c}{No} \\
        Traffic capacity (number of vehicles)  & 2,000\,--\,6,000 & 2,000\\
        \bottomrule
    \end{tabular}
\end{table*}

For the platooning use case, we assume that the vehicles, which consists of $N$ trucks, drive in a single-lane formation with a fixed inter-vehicle distance and coordinate their maneuvers. The first truck acts as ``platoon head'', the others as ``platoon member'' (PH and PM, respectively). The vehicles operate in sidelink mode\,4 when they are outside of coverage of an RSU. When cellular coverage is available, they can switch to mode\,3 or continue to remain in mode\,4 (Fig.~\ref{subfig:usecase}). %
In the chosen setup, an RSU operates as eNodeB.\footnote{An RSU can operate as UE- or eNodeB-type (3GPP TR 23.285 V14.2.0).} %
It stays connected to the infrastructure and is responsible for network-assisted V2I communication.

The platoon exchanges two types of messages (Table~\ref{tab:summary}): \emph{(i)}~all vehicles inside a platoon, including PH and PMs, transmit non-IP, V2V broadcast messages. %
Following ETSI standards, these messages have the type CAM, a variable size of 280-330\,bytes and are generated periodically~\cite{ETSI_CAM}. %
\emph{(ii)}~The PH exchanges IP-based, V2I unicast messages with an RSU and forwards the information from the RSU to the other trucks in the platoon by non-IP, V2V broadcast. %
These messages contain information about distant road conditions and traffic information; their size varies between 50 and 1,500\,bytes. %
We note that the scenario also contains other surrounding vehicles, which are not PM and periodically generate CAMs whose frequency is expected to vary between $\lambda = [1,10]$\,Hz. In the simulations, the nodes generate CAMs with a fixed period, which does not depend on the vehicle dynamics as in~\cite{ETSI_CAM}. Instead, we vary the CAM period in order to control the data traffic load. The generated traffic load in terms of payload size depends on the values of $\lambda$ and hence we consider normalized traffic load in our simulations. The mobility model for the platoon is adapted according to the ``ACC'' car following model in \texttt{SUMO} where the vehicles drive in accordance to the speed limits in Table~\ref{tab:params} while maintaining a minimum gap of 2.5\,m. 

\begin{table}[htb]
\centering
\caption{Overview of the simulated use case}
\label{tab:summary}
  \begin{tabular}{ll}
    \toprule
    Use case & V2X-based platooning\\

    Road type & Highway with tunnel \\

    Nodes & Platoon with 6 vehicles, RSU, \\
    & other vehicles with \\
    & 32\,vehicles/(km\,lane) \\

    Message types & Non-IP based CAM,\\
    & IP-based Alert\\

    Data exchange & V2I: RSU platoon head (PH)\\
    & V2V: platoon members (PM) \\

    Message & Unicast: RSU, infrastructure \\
    distribution & Unicast: RSU and PH \\ 
    & Broadcast: among PMs\\
    
    Interfaces & $U_{u}$ and PC-5 \\

    Mode switching& $U_{u}$ to PC-5 (mode\,3) \\
    & PC-5 sidelink mode\,3 to 4  \\
    Carrier frequency & V2I: $5.9$\,GHz, V2V: $5.9$ GHz \\
 \hline
  \end{tabular}
\end{table}

For the performance assessment of the ``in-platoon'' V2X communication for both modes, we consider two metrics for evaluation. The probability of message reception $P_{r}$ refers to the ratio of the number of messages successfully received ($N_{r}$) to the number of messages transmitted to the intended recipient  ($N_{t}$). The factors contributing to successful message exchange between UEs are: \emph{(i)}~the periodicity of CAMs and the reception time of alert messages from the RSU, \emph{(ii)}~the availability of resources in the selection window of SB-SPS and \emph{(iii)}~half-duplex constraints.\footnote{Due to the half-duplex constraints, a vehicle cannot receive a packet because it transmits its own packet in the same subframe.} The second metric, end-to-end (E2E) latency, measures the time taken for the transport of a CAM between the transmitting UE and the intended recipient UE. The E2E latency for alert messages is calculated as the time taken for the transport of message from the RSU to a PM via the PH. The E2E latency in both cases is affected by the resource allocation latency~\cite{mode} of the SB-SPS scheme.

\begin{figure*}[htb]
    \centering
    \subfloat[Probability of successful message reception $P_{r}$ (aggregated CAMs and alerts)]
    {\includegraphics[width=0.49\linewidth]{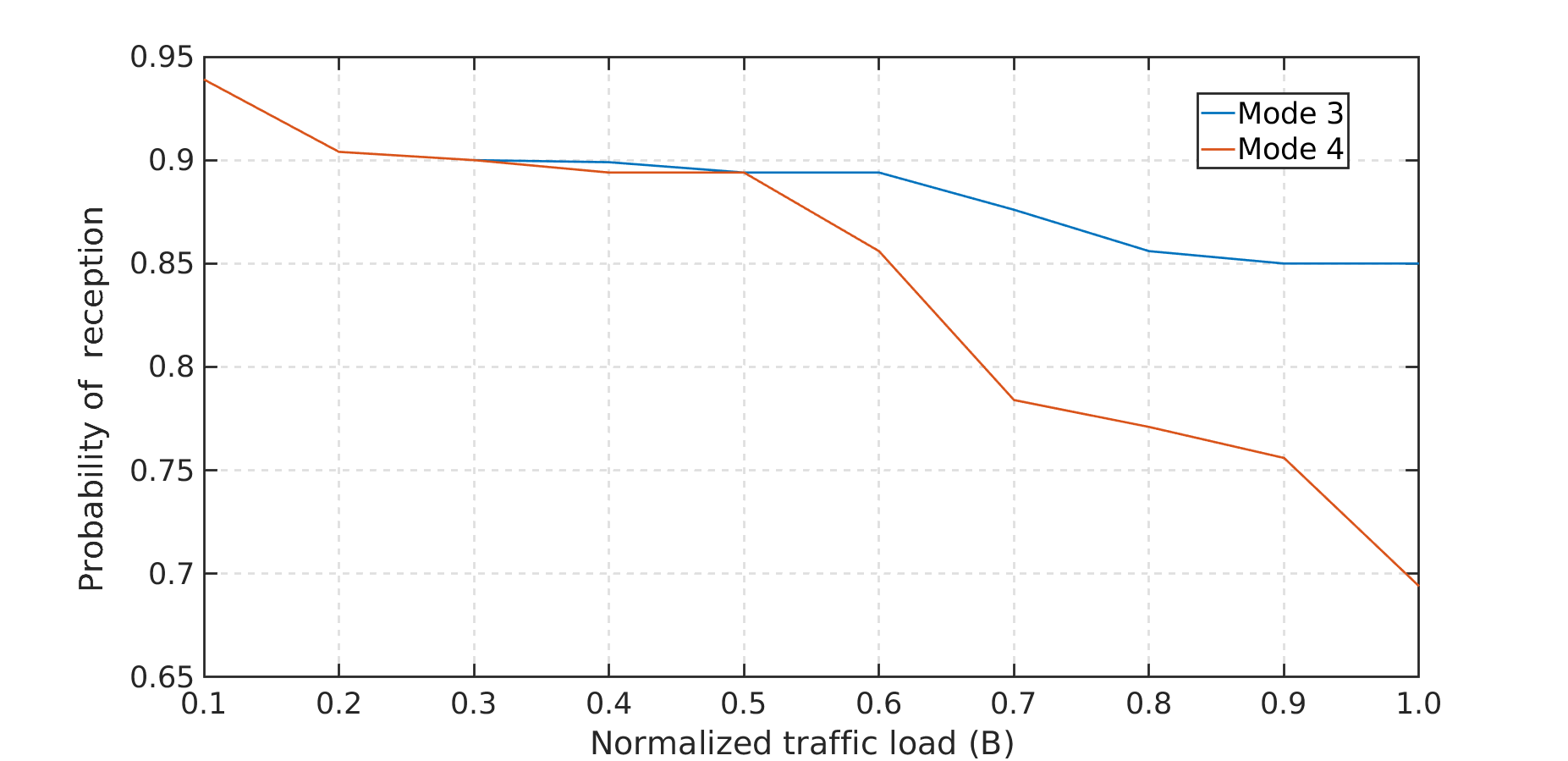}
    \label{subfig:reliability1}}
    \hfil
    \subfloat[End-to-end latency for CAMs and alerts]
    {\includegraphics[width=0.49\linewidth]{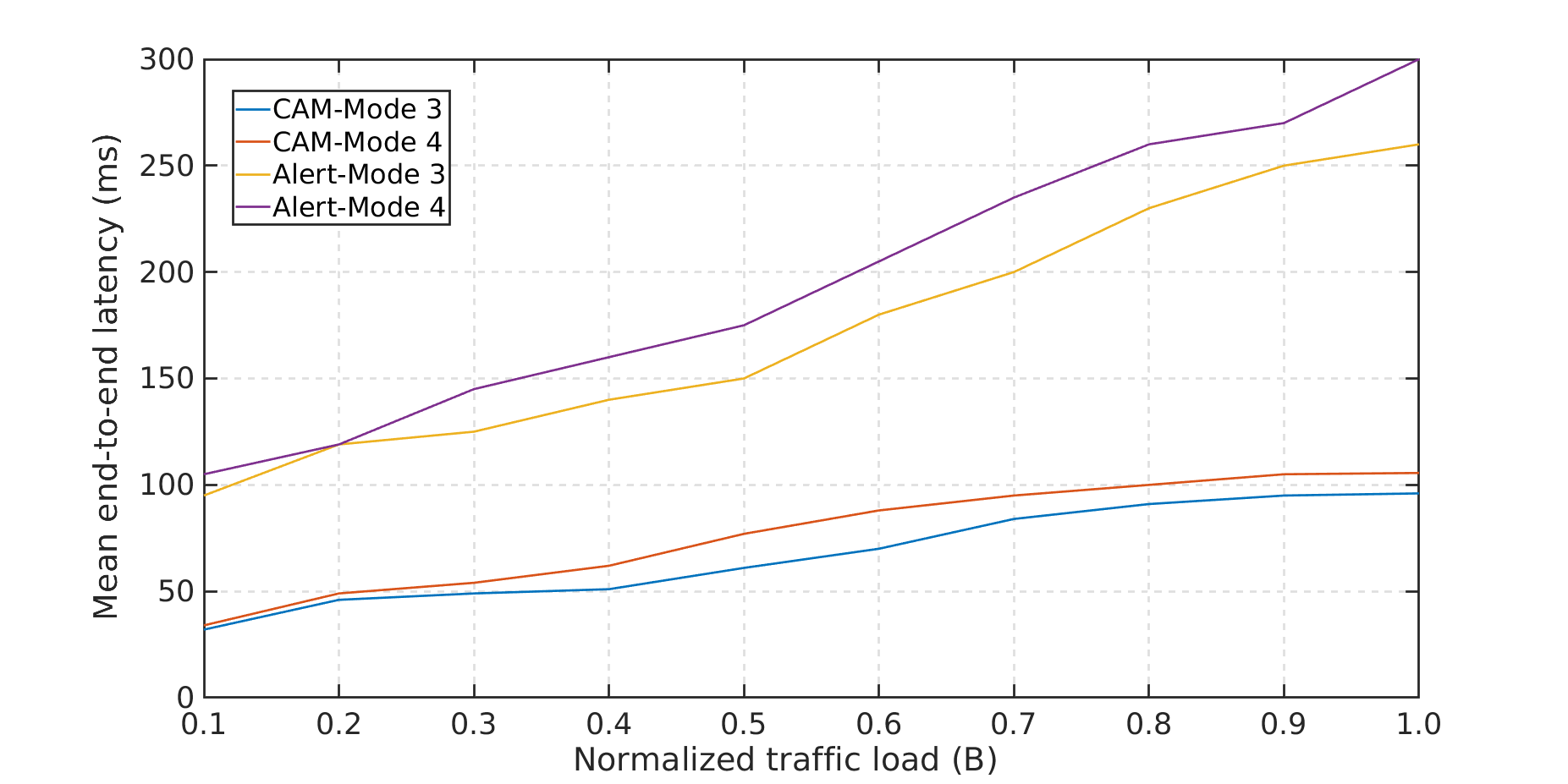}
    \label{subfig:latency}}
    \caption{Dissemination of periodic messages among vehicles in a motorway platoon: Comparison of mode\,3 \& 4} 
    \label{fig:modes_comparison}
\end{figure*}

\begin{figure}[h]
	 \centering
	 \includegraphics[width=1.05\linewidth]{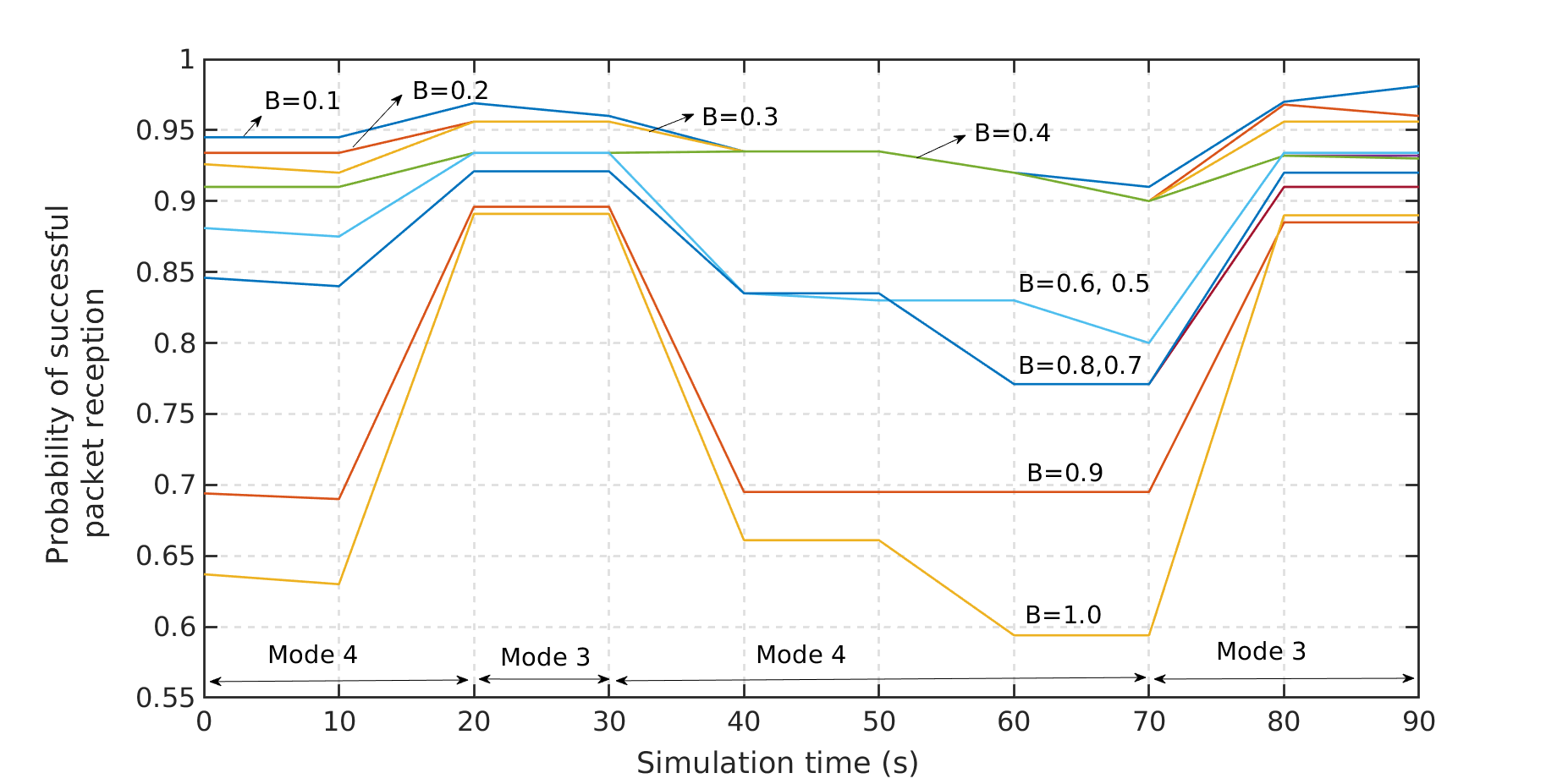}
	 \caption{Impact of mode switching on reliability}
	 \label{fig:mode_switch}
\end{figure}

For each of the above defined metrics, we make a comparison of the mode\,3 and mode\,4 performance as depicted in Fig.~\ref{fig:modes_comparison} and~\ref{fig:mode_switch}. 
In Fig.~\ref{subfig:reliability1}, we can observe that at lower traffic loads, $P_{r}$ is comparable for both modes. For small CAM generation frequencies, there is sufficient time for the CAMs to obtain resources and get transmitted. At medium-to-high traffic load, mode\,3 performs better than mode\,4 because the eNodeB is constantly aware of all the connected UEs, their traffic load and the resources utilized by them. Also, when an alert message interrupts a CAM transmission, the eNodeB employs a load balancing scheme to efficiently allocate resources for different message types. In mode\,4 when the CAM transmission of a vehicle is interrupted by another message of higher priority, i.e., alert, the UE has to immediately configure the sidelink grant and allocate resources for the higher priority message. If resources are still available in the selection window, CAMs can still get transmitted; otherwise they are lost. We stress that mode\,4 suffers from half duplex constraints where messages are lost at the receiving UE when a sender uses the same subframe to transmit its own CAM or alert message. This constraint does not depend on the distance between the transmitter and receiver but rather on the size of the subframe and the message rate of a vehicle. In case of mode\,4, the effect is more pronounced at higher traffic load.

In case of mode\,3, the eNodeB decides on the resource reservation and allocation based on the feedback it receives from the vehicle about the dynamically changing traffic load and the latency requirements. On the other hand, in mode\,4, the vehicles reserve their resources for several consecutive periodic message transmissions indicated by the $C_{resel}$. When the traffic load changes dynamically, several vehicles will compete for the same radio resources, which leads to multiple iterations of resource re-selections causing additional latency. %
However, with the vehicle density considered in our simulation scenario, we can see that the E2E latency for both CAM and alert messages meet the defined limits set by the standards~\cite{3gpp_usecases,ETSI_CAM} for both mode\,3 and mode\,4. 

Fig.~\ref{fig:mode_switch} illustrates the impact of mode switching on the reliability of packet transmission. We note that the PMs do not remain in the same mode throughout the simulation as in Fig.~\ref{fig:modes_comparison}. When the vehicles switch from mode\,4 to mode\,3, connection establishment and time synchronization with the \mbox{eNodeB} is faster than the vice versa process. As a result, the vehicles obtain resources within the expiry period of the message. %
This effect ensures that almost up to 90\,\% of the generated messages are successfully received. Depending on the location of the vehicle and the time at which it switches to mode\,4, it has to wait for the subsequent cycle to exchange synchronization subframes, which occurs at a period of 160\,ms, with other UEs. In situations where the traffic load is high and the synchronization gets slower, we can observe that ($P_{r}$) reduces to almost  60\,\%. This further supports the fact that it is always preferable for a vehicle to switch to mode\,3 if available.

\section{Conclusions}
\label{sec:concl}

We have presented \texttt{Artery-C}, an \texttt{OMNeT++}-based discrete event simulation framework for the assessment of Cellular V2X protocols and the evaluation of V2X application performance. \texttt{Artery-C} comprises control and user plane of Cellular V2X, implements components for every layer of the Cellular V2X protocol stack and realizes  up-/downlink and sidelink communication.
By seamless integration into the existing \texttt{Artery} framework, it facilitates the use of microscopic mobility models from SUMO, the simulation of the full C-ITS protocol stack including ad hoc networking, facilities, security and various other advanced features.
\texttt{Artery-C} meets the requirements for a comprehensive simulation framework related to software, Cellular V2X and timing. Using \texttt{Artery-C}, we also have presented performance results for V2X-based platooning in a highway scenario as a representative use case. These results demonstrate several capabilities of the simulator and validate technical key features.

\section{Acknowledgements}
This work was supported by the German Science Foundation (DFG) within the priority program Cooperatively Interacting Automobiles (CoInCar) (SPP 1835).
We would like to thank Mr. Raphael Riebl for insights about \texttt{OMNeT++} and \texttt{Artery}, Mr. Quentin Delooz and Ms. Julia Rainer for feedback and the developers of \texttt{SimuLTE} for providing their simulation environment to the open source community.




\end{document}